\documentclass[reprint,%
aps,prl,%
10pt,%
superscriptaddress,%
amsmath,amssymb,%
floatfix]{revtex4-2}

\usepackage{amsmath,amssymb,bm}
\usepackage{mathtools}
\usepackage{esvect}
\usepackage{booktabs}
\usepackage{multirow}
\usepackage{array}
\usepackage[svgnames]{xcolor}
\usepackage[colorlinks=true,%
linkcolor=MediumBlue,%
urlcolor=MediumBlue,%
citecolor=MediumBlue]{hyperref}
\usepackage{siunitx}
\DeclareSIUnit{\voltpermeter}{(\text{V}/\text{m})}

\renewcommand{\vec}[1]{\ensuremath{\mathbf{#1}}}

\newcommand{\Like}{\mathcal{L}}
\newcommand{\sign}[1]{\operatorname{sgn}\,#1}

\begin{document}
\title{A background-free optically levitated charge sensor}

\author{N.~Priel}
\email{nadavp@stanford.edu}
\affiliation{Department of Physics, Stanford University, Stanford, California 94305, USA}

\author{A.~Fieguth}
\affiliation{Department of Physics, Stanford University, Stanford, California 94305, USA}

\author{C.~P.~Blakemore}
\affiliation{Department of Physics, Stanford University, Stanford, California 94305, USA}

\author{E.~Hough}
\affiliation{Department of Physics, Stanford University, Stanford, California 94305, USA}

\author{A.~Kawasaki~\thanks{Now at AIST, Ibaraki, Japan}}
\thanks{Now at National Metrology Institute of Japan (NMIJ), National Institute of Advanced Industrial Science and Technology (AIST), 1-1-1 Umezono, Tsukuba, Ibaraki 305-8563, Japan}
\affiliation{Department of Physics, Stanford University, Stanford, California 94305, USA}
\affiliation{W.W.~Hansen Experimental Physics Laboratory, Stanford University, Stanford, California 94305, USA\looseness=-1}

\author{D.~Martin}
\affiliation{Department of Physics, Stanford University, Stanford, California 94305, USA}

\author{G.~Venugopalan} 
\affiliation{Department of Physics, Stanford University, Stanford, California 94305, USA}

\author{G.~Gratta}
\affiliation{Department of Physics, Stanford University, Stanford, California 94305, USA}
\affiliation{W.W.~Hansen Experimental Physics Laboratory, Stanford University, Stanford, California 94305, USA\looseness=-1}

\date{\today}

\begin{abstract}

Optically levitated macroscopic objects are a powerful tool in the field of force sensing, owing to high sensitivity, absolute force calibration, environmental isolation and the advanced degree of control over their dynamics that have been achieved. However, limitations arise from the spurious forces caused by electrical polarization effects that, even for nominally neutral objects, affect the force sensing because of the interaction of dipole moments with gradients of external electric fields.
In this paper we introduce a new technique to model and eliminate dipole moment interactions limiting the performance of sensors employing levitated objects. This process leads to the first noise-limited measurement with a sensitivity of $\SI{3.3e-5}{\elementarycharge}$. As a demonstration, this is applied to the search for unknown charges of a magnitude much below that of an electron or for exceedingly small unbalances between electron and proton charges. The absence of remaining systematic biases, enables true discovery experiments, with sensitivities that are expected to improve as the system noise is brought down to or beyond the quantum limit. As a by-product of the technique, the electromagnetic properties of the levitated objects can also be measured on an individual basis.

\end{abstract}

\maketitle

\section{Introduction}

Optical levitation of macroscopic objects in vacuum has recently drawn considerable attention due to numerous applications in the fields of sensing, quantum physics, and particle physics~\cite{Gonzalez-Ballestero:2021}. The versatility of this technique stems from the ability to measure and control the translation, rotation, charge state and dynamics of a macroscopic object with high precision~\cite{Magrini:2021,Tebbenjohanns:2021,Gieseler:2012, Jain:2016,Monteiro:2018, Rider:2019, Reimann:2018, Hoang:2016, Delord:2020, Ahn:2018,Stickler:2021}, where the thermal and electrical isolation of the levitated object from the environment make the low noise conditions possible. 

Work with levitated dielectric microspheres (MSs) with masses in the range  $\SIrange{0.1}{10}{\nano\gram}$ and force sensitivity $\simeq 10^{-18}~\SI{}{\newton\per\sqrt\hertz}$ can be applied to the investigation of phenomena beyond the Standard Model (BSM) of particle physics, including the search for a fifth-force at short range~\cite{Geraci:2010, Rider:2016, Blakemore:2021}, the breakdown of Coulomb’s law as a probe for a dark photon~\cite{Moore:2020}, high-frequency gravitational wave detection~\cite{Arvanitaki:2013, Aggarwal:2020}, and others~\cite{Carney:2021, Monteiro:2020, Afek:2021, Kawasaki:2018xak}. Each such endeavor is eventually expected to be limited by spurious electromagnetic interactions, ultimately limiting its sensitivity. Even for neutral levitated objects, such backgrounds are either a direct consequence of electromagnetic actuation and sensing~\cite{Moore:2014,Afek:2021a}, or an indirect effect arising from electric field gradients due to patch potentials~\cite{Rider:2016,Blakemore:2019,Blakemore:2021}.  In the broader context of experimental physics, systematic effects from residual interactions of non-uniform charge distributions have long plagued precision measurements~\cite{Everitt:2011,Buchman:2011,Armano:2017,Garret:2015,Lee:2020}.

In this work, we develop a model of those minute electromagnetic interactions with an optically trapped MS, and demonstrate a technique capable of identifying and eliminating the unwanted contributions.
This process enables noise-limited charge sensing at a level of $\SI{3.3e-5}{\elementarycharge}$ for macroscopic objects.

An obvious application of a macroscopic charge sensor with understood dynamics, is to probe the net neutrality of the sensor itself. The resulting value can be interpreted two ways with respect to the Standard Model of particle physics. Firstly, it can test the equality of magnitude of the proton and electron charges, complementing different techniques~\cite{Bressi:2011, Gallinaro:1977, Dylla:1973}.
Secondly, it can probe the existence of mini-charged particles (MCPs)~\cite{Dobroliubov:1990,Jaeckel:2010}. Such particles, not included in Standard Model, could help answer central questions in physics, such as how and why charge is quantized~\cite{Klein:1926, Dirac:1931}, and can also contribute to solve the Dark Matter puzzle~\cite{Pospelov:2008, Boehm:2004,McDermott:2010}. 

In more general terms, the new understanding of the electromagnetic dynamics presented here significantly enhances the power of BSM searches with optical levitation technique by allowing, for the first time, discovery potential.
In addition, this approach enables precision studies of the dielectric properties, e.g. polarizability, of levitated MSs.

\begin{figure*}[t]
    \includegraphics[width=1.5\columnwidth]{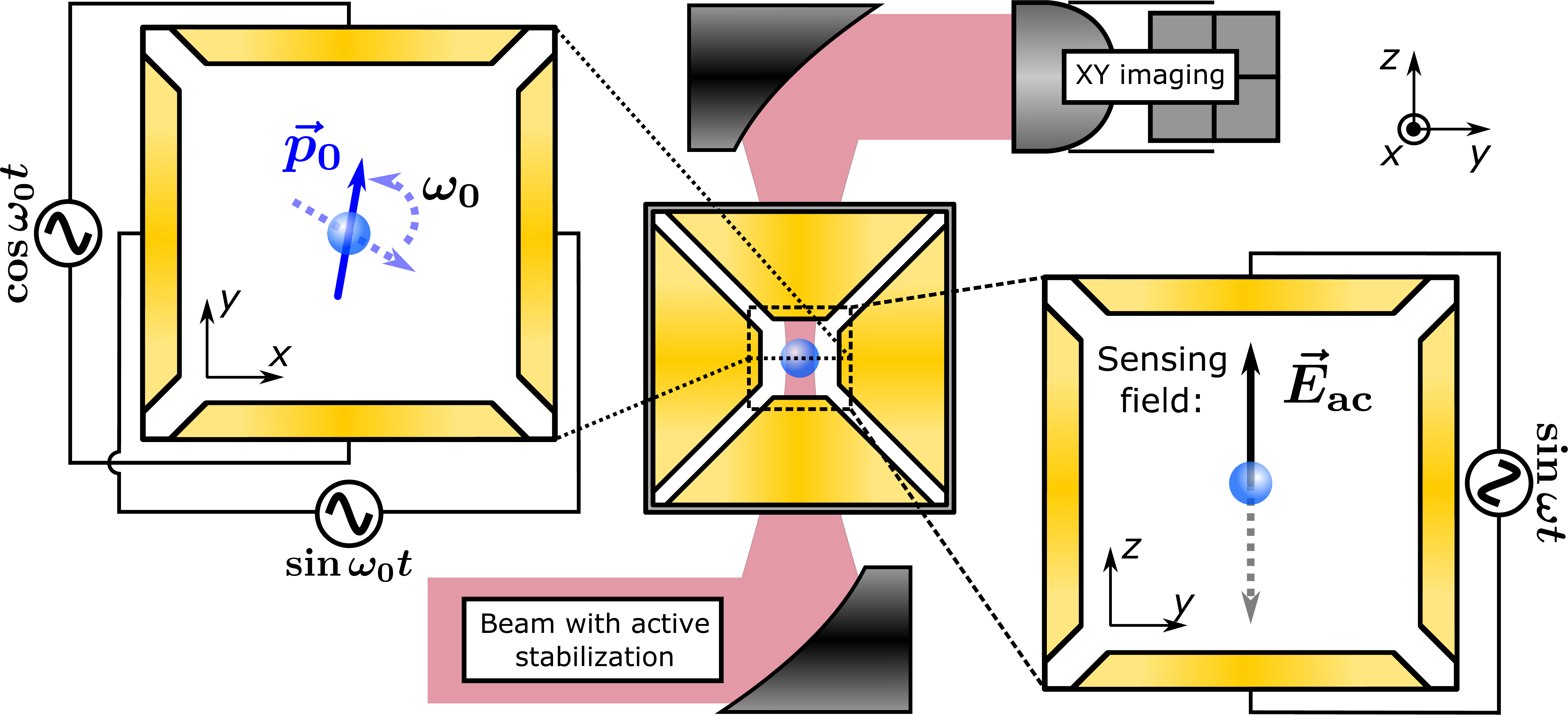}
    \caption{Schematic of the optical trap setup, with a MS at the center of the electrode cube structure. For clarity, the background and foreground electrodes have been omitted. A detail of the $xy$ plane is shown on the left, indicating the driven rotation of the dipole moment at $\omega_0 \sim 2 \pi (\SI{100}{\kilo\hertz})$ using the four horizontal electrodes, $x^{\pm}$ and $y^{\pm}$. A detail of the $yz$ plane is shown to the right, indicating the sensing field applied along the $z$-axis (vertical). This orientation of electric fields is given as a specific example, and other orientations (e.g. spinning in $yz$, sensing along $x$) are possible and have been implemented here.}
    \label{fig:scheme}
\end{figure*}

\section{Experimental setup and methodology}

The centerpieces of the experiment described here are silica~\cite{microparticles_gmbh} MSs, with a diameter of \SI{7.52\pm0.18}{\micro\meter}~\cite{Blakemore:2021a}, trapped through optical forces exerted by a \SI{1064}{\nano\meter} optical tweezer~\cite{Ashkin:1971} arrangement in vacuum. A schematic of the experimental setup is shown in Figure~\ref{fig:scheme}. The position of a MS in the horizontal ($xy$) plane is obtained by measuring the deflection of the transmitted portion of the trapping laser beam on a quadrant photodiode, while the vertical ($z$) position is obtained from the phase of the light retroreflected by the MS~\cite{Kawasaki:2020,Blakemore:2021a}. The optical trap is operated inside a vacuum chamber at  $\mathcal{O}$($10^{-6}~\SI{}{\hecto\pascal}$) and active feedback is employed to cool the three translational degrees of freedom of the MS. With this setup, a force sensitivity of $<10^{-16}~\SI{}{\newton\per\sqrt\hertz}$ has been achieved, which determines the ultimate sensitivity to any signal from the measurements presented here. 
The trap is closely surrounded by six identical electrodes forming a cubic cavity. Any configuration of voltages can be applied to the electrodes, producing specific electric field configurations at the location of the MS. The long term stability of the three-dimensional location is measured by an auxiliary imaging system, which is utilized to compensate for slow drifts in $z$ of the MS, maintaining the initial position with sub-\SI{}{\micro\meter} precision.
Two distinct optical traps, and  three MSs, have been used for the measurements described in this paper. The main difference between the traps being the separation between the faces of the electrodes, and hence, the size of the cubic region where the MS is trapped. The first setup, detailed in~\cite{Kawasaki:2020}, has the electrodes \SI{8}{\milli\meter} apart, while the second, described in~\cite{Blakemore:2021a}, has electrodes separated by \SI{4}{\milli\meter}.

\subsection{Force calibration and MS dipole stabilization}

MSs usually carry a net-charge when initially trapped, but their charge state can be altered in units of electron charge, by removing or adding individual electrons using UV photons from a xenon flash lamp. A $\mathcal{O}(100~\SI{}{\hertz})$ sinusoidal electric field with $\mathcal{O}(100~\SI{}{\volt})$ amplitude is applied with pairs of opposing electrodes, for the purpose of measuring the charge state and calibrating the MS response to applied forces. This can be achieved when the charge imbalance is of only a few $\SI{\pm1}{\elementarycharge}$, as the charge quantization becomes apparent with an exceedingly large S/N ratio~\cite{Kawasaki:2020, Moore:2014}. The MS is subsequently discharged to apparent net neutrality. 

During this calibration procedure, the response of the MS to a sinusoidal electric field applied to a single electrode, with all others grounded, is also measured. From this, the position of the MS (set by the trapping laser) relative to the centers of the three pairs of electrodes, can be estimated~\cite{Afek:2021bua}. This is done by comparing the ratio of the response to two opposing electrodes to the expected electric field ratio calculated with finite element analysis (FEA). This is distinct from the nominal force calibration performed earlier, wherein opposing electrodes are driven simultaneously to minimize gradients at the trap location.

Next, a rotating electrical field ($\vec{E}_{\mathrm{spin}}$) is applied (indicated by oscillating voltages applied to the $x^{\pm}$ and $y^{\pm}$ electrodes in Figure~\ref{fig:scheme}). This electric field applies a torque to the electric dipole moment that trapped MSs generally appear to carry~\cite{Rider:2016,Rider:2019,Blakemore:2020,Afek:2021b}, inducing the MS to rotate in the $xy$ plane with an angular velocity set by the frequency of the rotating field.  Typical electric fields have a frequency of ${\sim} \SI{100}{\kilo\hertz}$ and an amplitude of ${\sim}\SI{50}{\kilo\volt\per\meter}$. The plane of rotation can be set arbitrarily, and here rotation in $xy$ is used as a concrete example.

\subsection{Modeling electrostatic forces on a trapped MS}

Consider a trapped MS with a monopole charge $q$, and permanent dipole moment with magnitude $p_0$, subjected to an applied electric field $\vec{E}_{\mathrm{ac}}$ oscillating at $f_0$ along the $z$ axis (for rotation in $xy$), as well as a stray DC field $\vec{E}_{\mathrm{dc}}$ that is assumed to be constant in time. 

The application of the additional rotating field $\vec{E}_{\mathrm{spin}}$ confines the permanent dipole moment to a plane. Given that the rotation is much faster than $f_0$ and perpendicular to $\vec{E}_{\mathrm{ac}}$, the effective dipole moment is reduced, with a possible remaining effective value, $\vec{p}_{\mathrm{dc}}$, due to higher order electric moments or imperfect alignment between the electric fields and the (optical) axes of the trap.
The resulting force on the MS at $f_0$ can be separated into three distinct terms for individual consideration,
\begin{equation} \label{eq:background_model}
\vec{F} = \underbrace{q\vec{E}_{\mathrm{ac}}}_{\substack{\text{Monopole} }}\, +\,\underbrace{ \vec{p}_{\mathrm{dc}} \cdot \nabla \vec{E}_{\mathrm{ac}}}_{\substack{\text{Permanent dipole} }} \, + \, \underbrace{\vec{p}_{\mathrm{ac}} \cdot \nabla \vec{E}_{\mathrm{dc}}}_{\substack{\text{Induced dipole} }},
\end{equation}
\noindent with $\vec{p}_{\mathrm{dc}}$ being the time-averaged projection of the residual permanent dipole moment and $\vec{p}_{\mathrm{ac}}$ being the an oscillating dipole moment induced by the applied electric field. 

When biasing individual electrodes, the `Monopole' term in Equation~\ref{eq:background_model} is expected to have the opposite sign for opposing electrodes since $\vec{E}_{\mathrm{ac}}$ is always oriented toward (away from) the electrode producing the field when the applied voltage is positive (negative). 

The `Permanent dipole' term represents the interaction of the permanent dipole moment of the MS with electrical field gradients. This interaction has been the principal limitation for using levitated MS to search for new physics involving monopole charges~\cite{Moore:2014,Afek:2021c}. 

Unlike the monopole response, the contribution from the permanent dipole moment with a fixed orientation has the same sign for an electric field from each of two opposing electrodes. Hence a new differential measurement scheme in which the permanent dipole contribution is dynamically canceled can be introduced.  Let $\vec{F}^+$ ($\vec{F}^-$) be the force on the trapped MS due to an excitation field $\vec{E}_{\mathrm{ac}}$ sourced by the $z^+$ ($z^-$) electrodes, respectively, defined explicitly in the supplementary material. Let $F^{\pm}$ be the the projections of $\vec{F}^{\pm}$ along the $z$ axis where the sensing field is applied, so that the permanent dipole contribution to $F^{\pm}$ can be eliminated by constructing a combined response parameter $A$:
\begin{align} \label{eq:A_parameter}
A &\equiv F^+ - \eta F^- = 2 q E^+ + \left( p^+_{\mathrm{ac}} - \eta p^-_{\mathrm{ac}} \right) \frac{\partial E_{\mathrm{dc},z}}{\partial z}
\end{align}
\noindent with $\eta \equiv |\partial_z E^+| / |\partial_z E^-| = |E^+| / |E^-|$ a constant set by the alignment of the optical trap to the center of the electrode cube. The key point of this construction is the fact that $\sign{(E^+)}=-\sign{(E^-)}$ as na\"ively expected, but, importantly, $\sign{(\partial_z E^+)} = \sign{(\partial_z E^-)}$, which yields to the cancellation of the permanent dipole contribution. The value for $\eta$ used in our analysis is extracted from an FEA analysis of field gradients for the electrode geometry at the \emph{measured} position of the MS.

During a measurement sequence, the excitation sinusoidal field is sourced first from the $z^+$ electrode for \SI{10}{\second}. This is followed by a $\mathcal{O}$(1~\SI{}{second}) segment, where $\vec{E}_{\mathrm{ac}}$ is turned off, and $\vec{E}_{\mathrm{spin}}$ remains on, realigning the MS dipole moment to rotate within the $xy$ plane, correcting for any excursions that $\vec{E}_{\mathrm{ac}}$ may have introduced. Afterwards, the oscillating voltage is applied to the opposing electrode, $z^-$, for another \SI{10}{\second}, completing a single measurement sequence that is repeated for statistical robustness.

The `Induced dipole' term in Equation~\ref{eq:background_model} is not eliminated by the construction of the combined response parameter $A$. However, this contribution can be evaluated by introducing another construction, $B$, for the combined response at the second harmonic:
\begin{equation} \label{eq:B_parameter}
\begin{split}
B \equiv G^+ - \eta^2 G^- = -\frac{1}{2} \left( p^+_{\mathrm{ac}} - \eta p^-_{\mathrm{ac}} \right) \frac{\partial E^+}{\partial z}.
\end{split}
\end{equation}

Here $G^+$ ($G^-$) is the response amplitude at $2f_0$, when the $z^+$ ($z^-$) electrode is driven. It can be seen that $B$ is proportional to the remaining induced dipole component in $A$. By combining Equations~\ref{eq:A_parameter} and \ref{eq:B_parameter}, the monopole response can be isolated:
\begin{equation} \label{eq:monopole_final}
\begin{split}
2 q E^+ \simeq A + 2 B \frac{\left( \partial_z E_{\mathrm{dc},z} \right)}{\left( \partial_z E^+ \right)}
\end{split}
\end{equation}
\noindent where $\vec{E}_{\mathrm{dc}}$ is the stray electric field in the vicinity of the sphere. Assuming that $\vec{E}_{\mathrm{dc}}$ is originating from the stray voltages on the electrodes, measured to be on the order of \SI{10}{\milli\volt}, and taking into account the nominal values of $B$ (${\sim}\SI{5e-16}{\newton}$), it is found that the second term on the RHS of Equation~\ref{eq:monopole_final} is negligible throughout the current measurement. It has to be emphasized that the construction of the parameters $A$ and $B$ is advantageous not only to eliminate and describe the contributions of the dipole moments, but also in order to isolate those contributions and study them to learn about the dielectric properties of individual MSs under the conditions of the measurement.

\begin{figure*}[t]
    \includegraphics[width=2.\columnwidth]{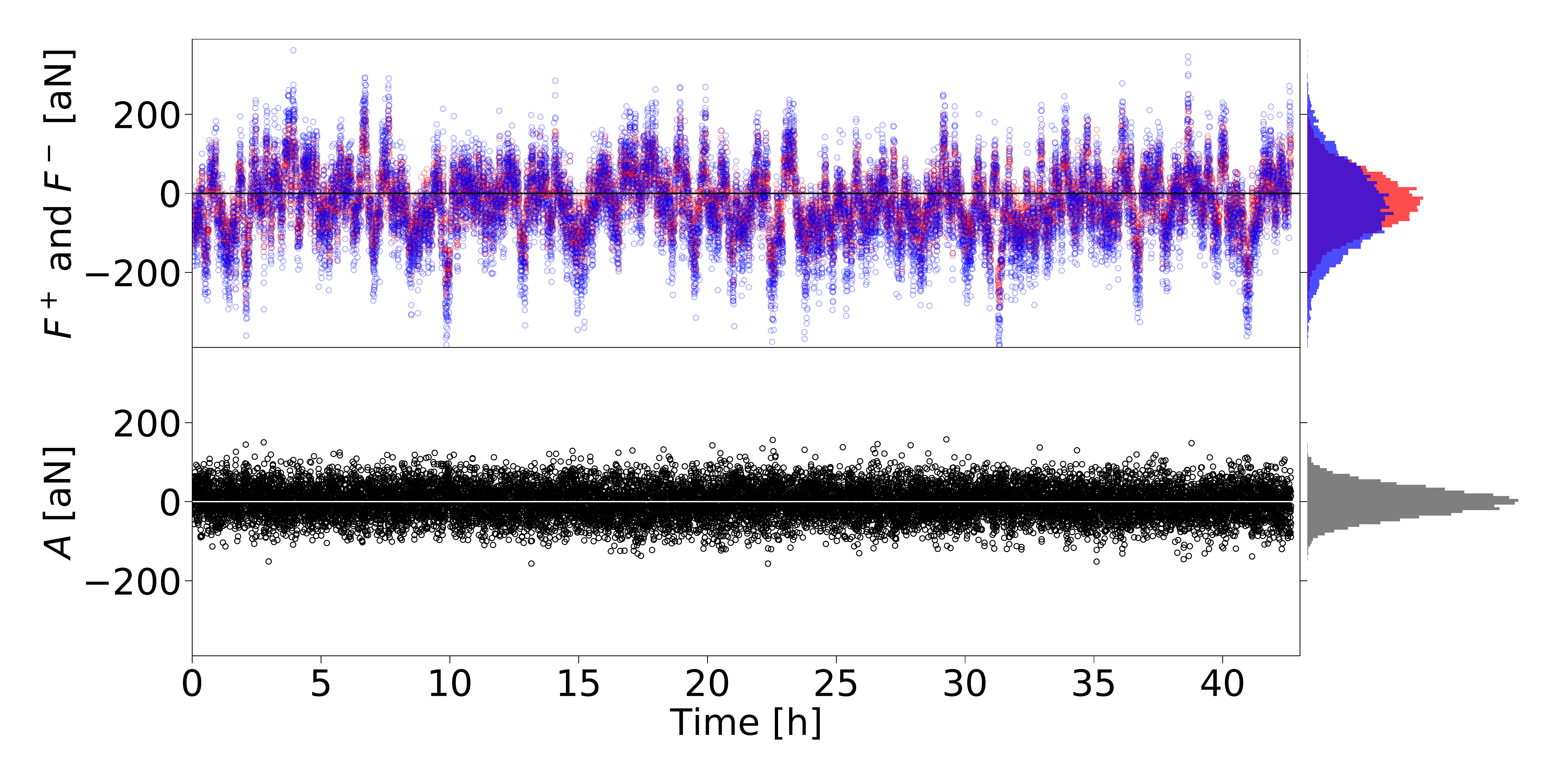}
    \caption{\textbf{Top:} In-phase responses, $F^+$ and $F^-$, of the MS to an oscillating electric field, $\vec{E}_{\mathrm{ac}}$, applied sequentially to the $z^+$ electrode (blue) and the $z^-$ electrode (red). Each point represents an average value of the response over a 10\,s data segment. The drifts in $F^+$ and $F^-$ are dominated by slow drifts in the permanent dipole moment. \textbf{Bottom:} Combined response parameter $A$ for the same data segments as in the top panel. Both panels also show equally-binned histograms projected on the response axis.} 
    \label{fig:force-vs-time}
\end{figure*}

\section{Results}

The measured response of a single MS spinning in the $xy$-plane, and driven at $f_0$ along $\hat{\vec{z}}$ by the $z^+$ and $z^-$ electrodes (one at a time), is shown in the top panel of Figure~\ref{fig:force-vs-time} (different orientations of electric fields were used for different MSs). The response is extracted by fitting a sine function to a bandpass filtered output from the MS's imaging system, where the phase for the fit is extracted from the digitized electrode voltage drive. For the same dataset, the behavior of the combined response parameter $A$ for the given responses $F^+$ and $F^-$ can be seen in the lower panel of Figure~\ref{fig:force-vs-time}. The cancellation of the residual permanent dipole effect is evident, and $A$ is purely limited by the noise floor.

\begin{figure}[!h]
    \includegraphics[width=1.\columnwidth]{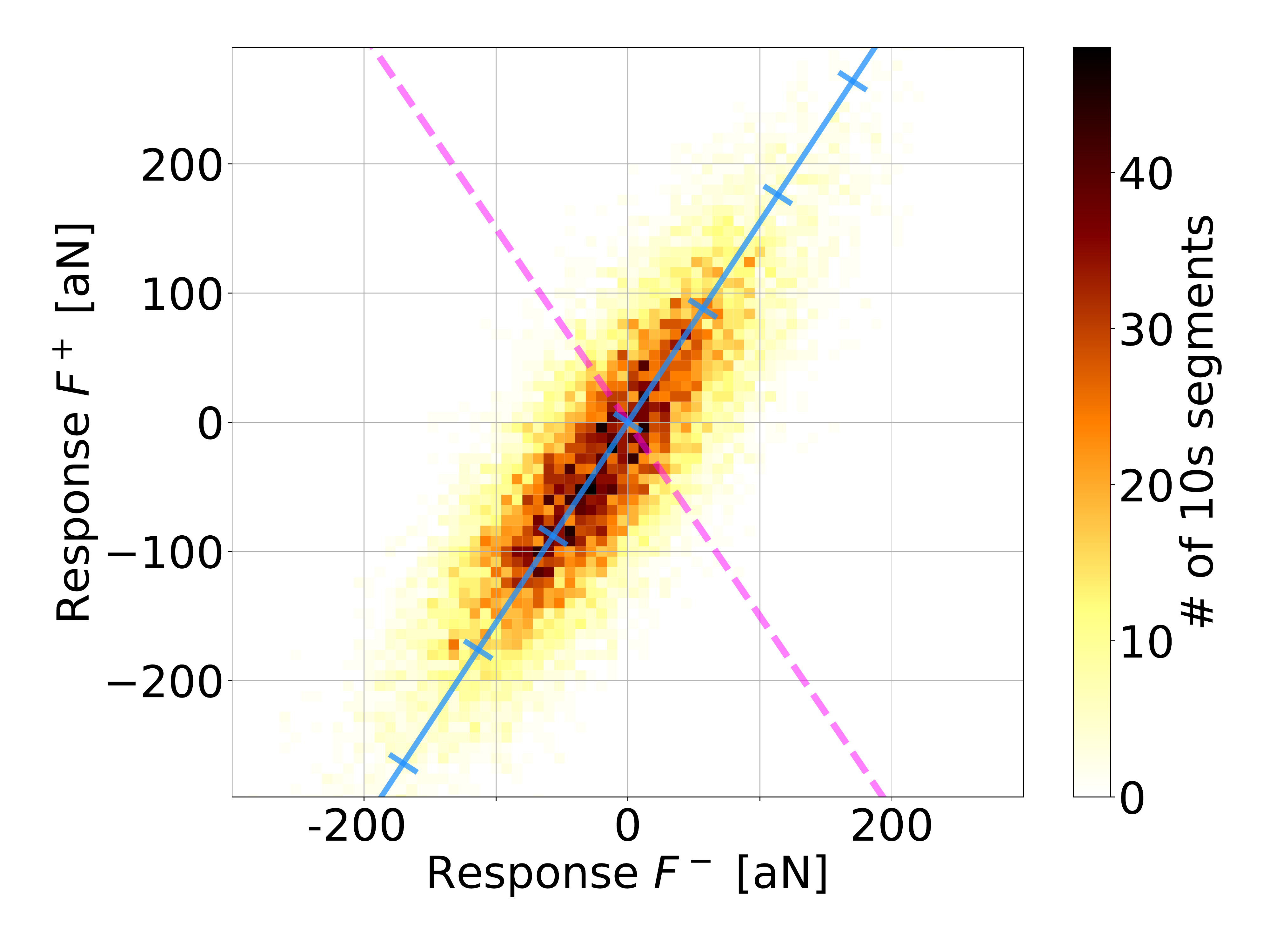}
    \caption{In-phase force responses, $F^+$ and $F^-$, of a MS, integrated over 10\,s. Voltages are applied sequentially to the $z^+$ and $z^-$ electrodes respectively. The data has been binned for illustration as indicated by the color scale. The dashed purple line represents an expected signal from the sphere carrying a monopole charge. The solid blue line represents the response expected from interaction of permanent dipole moments of varying magnitudes, with $\vec{E}_{\mathrm{ac}}$. Ticks on the blue line are separated by 25\,$e\cdot\mu m$.}
    \label{fig:scatter}
\end{figure}

Additionally, correlations between $F^+$ and $F^-$ in a data set collected over \SI{90}{\hour} are illustrated in Figure~\ref{fig:scatter}, where the superposed purple and blue line indicate expected signals due to the MS carrying a monopole charge and permanent dipole, respectively. One can see that most of the measured variation is consistent with a permanent dipole moment. This contribution is disentangled and eliminated by the combined response parameter $A$. The deviations from the origin along the major axis of the ellipse correspond to excursions of the rotating electric dipole moment from the $xy$-plane, as well as possible higher order electric moments.    Interestingly, $\left| \langle \vec{p}_{\mathrm{dc}} \rangle \right|$, the time-averaged value of the residual dipole moment along $z$, can be estimated by utilizing $F^+ + |E^+/E^-|F^- \simeq 2 p_{dc} \cdot \nabla E^+$. For the data set presented in Figures~\ref{fig:force-vs-time} and \ref{fig:scatter}, this is consistent with a $\left| \langle \vec{p}_{\mathrm{dc}} \rangle \right| \lesssim \SI{20}{\elementarycharge \micro\meter}$. The exact value of this contribution in individual \SI{10}{\second} data segments varies over a few hours.  The value is, nevertheless, an order of magnitude smaller than the typical electric dipole moment of the MS, that is the relevant quantity for non-spinning MSs.

In a similar fashion, $G^+$ and $G^-$ can be used to estimate $p^+_{\mathrm{ac}}$ and $p^-_{\mathrm{ac}}$ to be \SI{214}{\elementarycharge \micro\meter} and \SI{318}{\elementarycharge \micro\meter}, respectively. The dominant sources of $p^{\pm}_{\mathrm{ac}}$ are the inherent polarizability of the MS, expected to be $\mathcal{O}(\SI{1e-2}{\elementarycharge \micro\meter \per \voltpermeter})$, and the oscillations of the dipole moment normal to the plane of rotation driven by $\vec{E}_{\mathrm{ac}}$. Those two sources will be investigated in future work, possibly by inducing a change in the polarizability due to faster rotation~\cite{Hummer:2020}, and/or by applying $\vec{E}_{\mathrm{ac}}$ in the plane of rotation.

The datasets used for this analysis were taken with three different MSs, at distinct electric field amplitudes, frequencies, spin axes, and overall integration times, as described in Table~\ref{tab:feild_configuration}. The charge sensitivity of each MS in the table is estimated by fitting a Gaussian to the distribution of $A$ (lower-right panel in Figure~\ref{fig:force-vs-time}). The mean value of each fit is compatible with zero.

\begin{table*}[t!]
\centering
\begin{center}
\caption{Electric field configuration, where the value of the electric field for analysis is derived from the FEA. The field strength can be approximated as $|\vec{E}_{x_i}| \sim 0.65 \Delta V / \Delta x_i$, with $\Delta x_i$ the relevant electrode separation.}
\label{tab:feild_configuration}
\begin{tabular}{*{7}{c}}
    \hline
    \hline
         & \multicolumn{3}{c}{\textbf{Oscillating field}}&  \\
        \cmidrule(lr){2-4}
        \textbf{MS} & \textbf{Voltage ($\Delta V$)} & \textbf{Frequency ($f_0$)} & \textbf{Axis} &  \textbf{Electrode separation} & \textbf{Integration time} & \textbf{Charge sensitivity} \\
    \hline
        1 & 20\,V & 71\,Hz & x & $\Delta x = 8$\,mm & 27\,h & $\SI{4.5e-4}{\elementarycharge}$ \\
        2 & 200\,V & 71\,Hz & y & $\Delta y = 8$\,mm & 28\,h & $\SI{7.7e-5}{\elementarycharge}$ \\
        3 & 200\,V & 139\,Hz & z & $\Delta z = 4$\,mm & 92\,h & $\SI{3.9e-5}{\elementarycharge}$\\

    \hline
    \hline
\end{tabular}
\end{center}
\end{table*}

In order to combine the datasets of the three different MSs, the combined likelihood function is constructed:
\begin{equation} \label{eq:likelihood}
\Like (q) = \prod_{ij} \frac{1}{\sqrt{2 \pi \sigma_{ij}^2}} \exp \bigg\{ \frac{-[A_{ij}/2 E_i^{{+}} - q)]^2}{2 \sigma_{ij}^2} \bigg\}.
\end{equation}

\begin{figure}[!b]
    \includegraphics[width=1.\columnwidth]{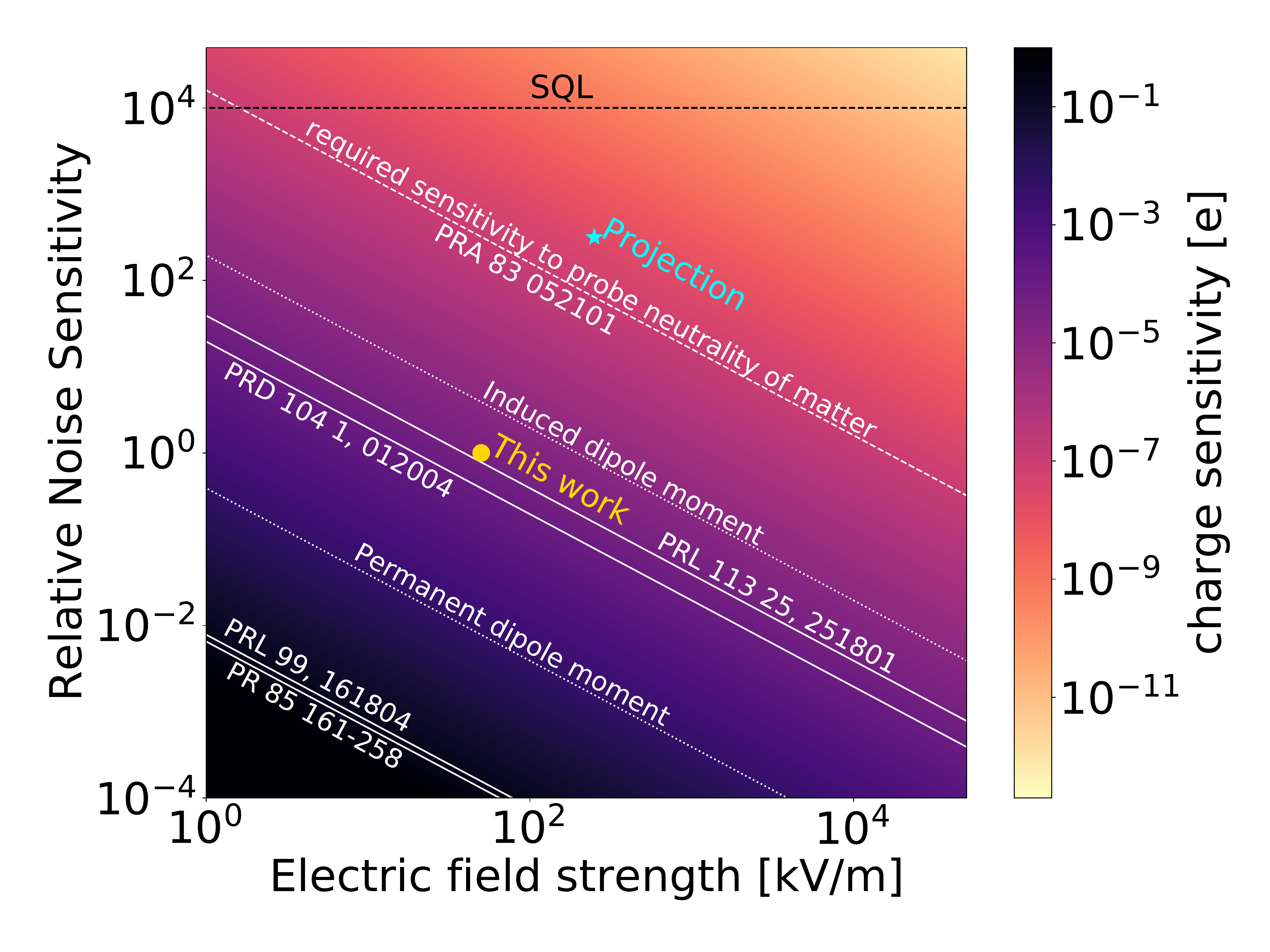}
    \caption{Charge sensitivity (color coding) as a function of electric field strength and noise sensitivity relative to the presented result (gold dot). Previous experiments~\cite{Kim:2007,Marinelli:1982,Moore:2014,Afek:2021a} have been added as contours of charge sensitivity. In addition a typical signal arising from a permanent dipole moment of the order of $\mathcal{O}$(100 \SI{}{\elementarycharge\micro\meter}) has been added. While the particular magnitude is specific to the gradients in the presented setup, signals of this order have been observed in all techniques sensitive to the dipole moment of a levitated MS~\cite{Moore:2014,Afek:2021,Afek:2021b}. Based on the presented setup and estimated stray fields sourced by contact potentials of $\mathcal{O}$(\SI{10}{\milli\volt}), an expected signal from the induced dipole at the driving frequency has also been added as a charge sensitivity contour. Once this signal has been properly measured, as the contribution was observed to be negligible in the current work, nothing prohibits further extension of the reach of the technique. A future projection of the presented experiment is added, where an improvement of the relative noise sensitivity by a factor of 300 and an increase of the electric field by a factor of five have been assumed. A measurement with this charge sensitivity could easily improve on the existing limits on the measurement of neutrality of matter (dashed line) for the given MS. A possible future limitation by the standard quantum noise limit (SQL) is indicated and calculated from the radiation pressure using the same sized MS.}
    \label{fig:charge_sens}
\end{figure}

Here, the index $i$ is running over the three MSs, and the index $j$ is running over all measured $A$s for a given sphere. The standard deviation, $\sigma_{ij}$, is estimated using the same fitting algorithm used to extract $A$, but on a sideband frequency separated from $f_0$ by \SI{1}{\hertz}, sufficient to exclude any contribution from the component at $f_0$. The combined sensitivity ($\pm \, 1\sigma$) of the measurement is $\SI{3.3e-5}{\elementarycharge}$ and the combined result is compatible with the no-monopole charge hypothesis. This represents the first high sensitivity, background-free search for monopole interactions using optically levitated microspheres. The absence of any monopole signal can be translated into a limit on the abundance of MCP bound to matter that is competitive with existing state-of-the art measurements~\cite{Moore:2014,Afek:2021c}, and complementary to other searches for MCPs~\cite{Prinz:1998, CMS:2013, Davidson:1991, Badertscher:2007, Gninenko:2007, Vinyoles:2016, Davidson:2000, Kamiokande-II:1991, Majorana:2018, SuperCDMS:2020, Magill:2018}. Alternatively, the data can be used to set a limit on the neutrality of matter, although with a lower sensitivity compared to~\cite{Bressi:2011, Gallinaro:1977, Dylla:1973}. Details of these two interpretations are presented in the Supplementary Material.

\section{Future prospects}

The technique illustrated here is only limited by the signal-to-noise ratio (SNR) of the current setup. Figure~\ref{fig:charge_sens} summarizes possible charge sensitivity improvements as a function of the two parameters responsible for the SNR: the strength of $\vec{E}_{\mathrm{ac}}$ and the relative noise sensitivity, normalized to the current setup. The electric field strength is currently limited by the existing instruments and can be increased with careful upgrade of the electrode configuration and vacuum feedthroughs. The relative noise sensitivity can be increased by extending the measurement time, assuming the integrability of noise, and by reducing the inherent force noise of the system. Sources of noise such as power and pointing fluctuations of the trapping laser can be mitigated in future iterations, where similar setups have already shown improvements by a factor of 10-100~\cite{Kawasaki:2020,Monteiro:2020a}.

Using parameters that are achievable in the near-future, namely an electric field strength of \SI{150}{\kilo\volt\per\meter}, an integration time of $10^{3}~\SI{}{hour}$ and a force noise of $10^{-18}~\SI{}{\newton\per\sqrt\hertz}$, a charge sensitivity of ${\sim}\SI{4e-8}{\elementarycharge}$ is expected. Such a charge sensitivity would be sufficient to improve the sensitivity to neutrality of matter by about an order of magnitude, as shown in Figure~\ref{fig:charge_sens}. The sensitivity could be further enhanced by using larger MS, assuming no degradation of noise performance.

Figure~\ref{fig:charge_sens} also shows the relative noise sensitivity of the standard quantum limit (SQL), at which level a charge sensitivity of $\mathcal{O}(10^{-11}~\SI{}{\elementarycharge})$ would be achievable.

\section{Conclusions}

We have presented the first background-free search for charges much smaller than the electron charge $\SI{}{\elementarycharge}$ with an optically levitated macroscopic sensor. By modelling and eliminating contributions due to the permanent dipole moment interacting with electric field gradients a force-noise-limited charge sensitivity of $\SI{3.3e-5}{\elementarycharge}$ was achieved. Measured data from three different microspheres has been analyzed using a profile likelihood method to explore the parameter space for mini-charged particles and the overall neutrality of matter. No monopole signal excess has been found in the present iteration of the experiment. With modest improvements of the setup, a future iteration of the experiment can be competitive with, or improve upon, the current leading experiments probing the neutrality of matter and searching for fifth-forces.
An exciting future prospect lies in applying this technique to perform metrology on the electromagnetic properties of levitated objects, such as their polarizability and permittivity, \textit{in situ}. This will enable a more detailed understanding of the dynamics of such objects, necessary in order to push the boundaries of the technology towards the quantum limit.

\section{Acknowledgements}
This work was supported by the National Science Foundation under grant No. PHY2108244, the Office of Naval Research under Grant No. N00014-18-1-2409 and the Heising-Simons Foundation. N.P. is supported, in part, by a grant of the Koret Foundation. C.P.B. acknowledges the partial support of a Gerald J. Lieberman Fellowship of Stanford University. A.K. was partial supported by a William M. and Jane D. Fairbank Postdoctoral Fellowship of Stanford University. We acknowledge regular discussions on the physics of trapped microspheres with the group of Prof. D. Moore at Yale. 

\bibliography{minicharge}

\end{document}


\title{Supplementary materials for: A background-free optically levitated charge sensor}

\author{N.~Priel}
\affiliation{Department of Physics, Stanford University, Stanford, California 94305, USA}

\author{A.~Fieguth}
\affiliation{Department of Physics, Stanford University, Stanford, California 94305, USA}

\author{C.~Blakemore}
\affiliation{Department of Physics, Stanford University, Stanford, California 94305, USA}

\author{E.~Hough}
\affiliation{Department of Physics, Stanford University, Stanford, California 94305, USA}

\author{A.~Kawasaki~\thanks{Now at AIST, Ibaraki, Japan}}
\affiliation{Department of Physics, Stanford University, Stanford, California 94305, USA}
\affiliation{W.W.~Hansen Experimental Physics Laboratory, Stanford University, Stanford, California 94305, USA\looseness=-1}

\author{D.~Martin}
\affiliation{Department of Physics, Stanford University, Stanford, California 94305, USA}

\author{G.~Venugopalan} 
\affiliation{Department of Physics, Stanford University, Stanford, California 94305, USA}

\author{G.~Gratta}
\affiliation{Department of Physics, Stanford University, Stanford, California 94305, USA}
\affiliation{W.W.~Hansen Experimental Physics Laboratory, Stanford University, Stanford, California 94305, USA\looseness=-1}

\date{\today}

\maketitle

\section*{Signal and Background Model}

Consider a trapped MS with a monopole charge $q$, permanent dipole moment  with magnitude $p_0$, and electric susceptibility $\chi$, subjected to an applied electric field $\vec{E}_{\mathrm{ac}}$ oscillating at $f_0$, as well as a stray dc field $\vec{E}_{\mathrm{dc}}$ that is assumed to be constant in time. Given that the permanent dipole moment is driven to rotate at a frequency $f_\text{spin}$ much larger than $f_0$, much of the effect of $\vec{p}_0$ can be averaged away. Higher order electric moments or non-zero, time-averaged projections of the spinning dipole along a measurement axis can contribute to an apparent residual permanent dipole moment $\vec{p}_{\mathrm{dc}}$. 

The force on a monopole charge is trivially $\vec{F}_q = q (\vec{E}_{\mathrm{dc}} + \vec{E}_{\mathrm{ac}})$, and thus, ignoring the constant, only has a term at the fundamental driving frequency $f_0$. The force on a dipole within a non-uniform electric field can be calculated as,
\begin{equation} 
\begin{split}
\vec{F}_{\vec{p}} &= - \nabla U_{\vec{p}} = \nabla (\vec{p} \cdot \vec{E}) \\
&= (\vec{p} \cdot \nabla)\vec{E} + (\vec{E} \cdot \nabla)\vec{p} \\
&\hspace*{1.3cm} + \vec{p} \times (\nabla \times \vec{E}) + \vec{E} \times (\nabla \times \vec{p}) \\
&= (\vec{p} \cdot \nabla)\vec{E},
\end{split}
\end{equation} where the first three terms of the expanded vector identity vanish since $\nabla \times \vec{E}=0$ for non relativistic fields, and $\nabla \cdot \vec{p} = \nabla \times \vec{p} = 0$ as the dipole moment is not a function of position and is assumed to have a physical extent smaller than the rate of change of the local gradients. The final result is then often written simply as $\vec{p} \cdot \nabla \vec{E}$. The dipole moment relevant to this calculation includes both the residual projection of the spinning dipole, as well as a term induced by the applied oscillating field, which yield $\vec{p} = \vec{p}_{\mathrm{dc}} + \vec{p}_{\mathrm{ac}}$. The total force on the MS is then,
\begin{equation} \label{eq:total_force}
\vec{F} = \underbrace{ q \vec{E}_{\mathrm{ac}} + \vec{p}_{\mathrm{dc}} \cdot \nabla \vec{E}_{\mathrm{ac}} + \vec{p}_{\mathrm{ac}} \cdot \nabla \vec{E}_{\mathrm{dc}} }_{ \text{Terms at } f_0 } + \underbrace{ \vec{p}_{\mathrm{ac}} \cdot \nabla \vec{E}_{\mathrm{ac}} }_{ \text{Term at } 2f_0 },
\end{equation}
\noindent

\subsection*{MS Response At The Fundamental Frequency}

The resulting force on a trapped MS at the drive's fundamental frequency $f_0$ can be separated into three distinct terms for individual consideration,
\begin{equation} \label{eq:background_model}
\vec{F} = \underbrace{q\vec{E}_{\mathrm{ac}}}_{\substack{\circled{1} \\ \text{Monopole} }}\, +\,\underbrace{ \vec{p}_{\mathrm{dc}} \cdot \nabla \vec{E}_{\mathrm{ac}}}_{\substack{\circled{2} \\ \text{Permanent dipole} }} \, + \, \underbrace{\vec{p}_{\mathrm{ac}} \cdot \nabla \vec{E}_{\mathrm{dc}}}_{\substack{\circled{3} \\ \text{Induced dipole} }},
\end{equation}
\noindent with $\vec{p}_{\mathrm{dc}}$ being the time-averaged projection of the permanent dipole moment or the component(s) of a higher order charge distribution (or a linear combination), and with the volume integral $\vec{p}_{\mathrm{ac}} = \int \epsilon_0 \chi \vec{E}_{\mathrm{ac}} \, dV$ being the dipole moment induced by the applied electric field. The stray field $\vec{E}_{\mathrm{dc}}$ would also induce a dipole moment by the same mechanism, thus contributing to $\vec{p}_{\mathrm{dc}}$.

For concreteness, we will assume a sensing field $\vec{E}_{\mathrm{ac}} = E \sin{(2 \pi f_0 t)} \, \vec{\hat{z}}$, where $E = E(\vec{x},V_{\alpha})$ is a scalar function of spatial coordinates within the trapping region and has been modeled with FEA for any combination of electrostatic drive voltages $V_{\alpha}$ applied to the $\alpha$ electrodes. The monopole response is then given by:
\begin{equation} \label{eq:monopole_response}
\circled{1} = q \vec{E}_{\mathrm{ac}} = q E_0(V_{\alpha}) \sin{(2 \pi f_0 t)} \, \vec{\hat{z}},
\end{equation}
\noindent where $E_0(V_{\alpha}) = E(\vec{x}_0, V_{\alpha})$ is the electric field evaluated at the position of the optical trap, $\vec{x}_0$.

For the second term, the $i$-th component of the directional derivative of $\vec{E}_{\mathrm{ac}}$ along $\vec{p}_{\mathrm{dc}}$ is given by,
\begin{equation} \label{eq:permanent_dipole_response}
\begin{split}
\circled{2} &= (\vec{p}_{\mathrm{dc}} \cdot \nabla E_{\mathrm{ac},i}) \, \vec{\hat{x}}^i \\
&= \left( p_{\mathrm{dc},x} \frac{\partial }{\partial x} + p_{\mathrm{dc},y} \frac{\partial }{\partial y} + p_{\mathrm{dc},z} \frac{\partial }{\partial z} \right) E_{\mathrm{ac},z} \, \vec{\hat{z}},
\end{split}
\end{equation}
\noindent where $E_{\mathrm{ac},x} = E_{\mathrm{ac},y} = \nabla E_{\mathrm{ac},x} = \nabla E_{\mathrm{ac},y} = 0$ by construction. Imperfect alignment both of the MS relative to the center of the electrode cube, as well as between the electrodes themselves, can violate this assumption. An FEA together with the measured position of the MS suggest that residual fields and gradients in the $x$ and $y$ directions contribute at the sub-percent level. Indeed, these would also induce the MS to respond in $x$ and $y$, whereas the measurement is performed in $z$ for this particular configuration of fields.

The term \circled{2} can be further simplified by two simultaneously observations. Firstly, since the dipole moment is being driven to rotate at high frequency f$_\text{spin}$ in the $xy$-plane, $p_{\mathrm{dc},x}$ and $p_{\mathrm{dc},y}$ should average to zero over one cycle of the sensing field's oscillation (since $f_\text{spin} \gg$ $f_0$). Furthermore, FEA suggests that $\partial_x E_{\mathrm{ac},z} \sim \partial_y E_{\mathrm{ac},z} \ll \partial_z E_{\mathrm{ac},z}$ at the measured position of the trap, by design of the electrode structure (where $\partial_{x}$ is shorthand for $\partial / \partial x$). This leaves us with a single term:
\begin{equation} \label{eq:permanent_dipole_response_simplified}
\circled{2} \simeq p_{\mathrm{dc},z} \frac{\partial E_{\mathrm{ac},z}}{\partial z} \, \vec{\hat{z}}.
\end{equation}

Continuing with \circled{3}, note that the induced dipole $\vec{p}_{\mathrm{ac}}$ will necessarily be aligned with the applied sensing field so that $\vec{p}_{\mathrm{ac}} = p_{\mathrm{ac}} \, \vec{\hat{z}}$ so that the term is given by,
\begin{equation} \label{eq:induced_dipole_response}
\begin{split}
\circled{3} &= (\vec{p}_{\mathrm{ac}} \cdot \nabla E_{\mathrm{dc},i}) \, \vec{\hat{x}}^i \\
&= p_{\mathrm{ac}} \frac{\partial}{\partial z} \left( E_{\mathrm{dc},x} \, \vec{\hat{x}} + E_{\mathrm{dc},y} \, \vec{\hat{y}} + E_{\mathrm{dc},z} \, \vec{\hat{z}} \right).
\end{split}
\end{equation}

For the current apparatus, we have no direct means of measuring $\vec{E}_{\mathrm{dc}}$, although it can be constrained by examining the response of the MS at $2 f_0$. For the work presented here, terms not along the measurement axis, $z$, have been ignored. Indeed, as the analysis will show, the induced dipole background term is below the noise floor of the MS's response at $f_0$.

Altogether, the expected force on a MS in response to a sensing field oscillating at $f_0$, considering only the response along the measurement axis $\vec{\hat{z}}$ is thus given by,
\begin{equation} \label{eq:final_model}
\vec{F} \cdot \vec{\hat{z}} = q E_{\mathrm{ac},z} + p_{\mathrm{dc},z} \frac{\partial E_{\mathrm{ac},z}}{\partial z} + p_{\mathrm{ac}} \frac{\partial E_{\mathrm{dc},z}}{\partial z},
\end{equation}
\noindent with $E_{\mathrm{ac},z} = E \sin{(2 \pi f_0 t)}$ as before, and where it is implicitly assumed that the field and field gradient terms are evaluated at the position of the optical trap, $x_0$.

\subsection*{Iterative Measurements and the \ensuremath{A} Parameter}

In nearly all previous searches for minute monopole charges with levitated test masses, the response represented by Equation~\ref{eq:final_model} was measured directly, for a test mass at apparent net neutrality (equal numbers of $p$ and $e^-$), and subsequently used to constrain possible values of $q \neq 0$. The observed charge sensitivity can be then translated to parameter spaces of Beyond the Standard Model theories, under certain assumptions. Importantly in this approach, any measured response incompatible with zero, for an otherwise electrically neutral test mass, could be explained directly by both permanent and induced dipole moments, as demonstrated by Equation~\ref{eq:final_model}. Historically, this has eliminated any discovery potential, and it would be impossible to deconvolve the monopole response of MCPs from the permanent and induced dipole backgrounds, even when the dipolar effects are minimized through techniques like electrically driven rotation.

Consider instead, two distinct measurement configurations: one with the electric field sourced from the $z^+$ electrode, and another with the electric field sourced from the $z^-$ electrode, generating forces $\vec{F}^+$ and $\vec{F}^-$, respectively. For both configurations, we will assume a sinusoidal driving voltage with identical amplitude and phase, such that the monopole response has a sign flip between the configurations. Explicitly,
\begin{equation} \label{eq:plus_and_minus}
\frac{\vec{F}^{\pm} \cdot \vec{\hat{z}}}{ \sin{(2 \pi f_0 t)} } = \left( q E^{\pm} + p_{\mathrm{dc},z} \frac{\partial E^{\pm} }{\partial z} + p^{\pm}_{\mathrm{ac}} \frac{\partial E_{\mathrm{dc},z}}{\partial z} \right),
\end{equation}
\noindent where $E^{\pm}$ are sourced by an \emph{identical} driving voltage, so that, relative to the fixed coordinate system established by the electrodes, $E^+ = \eta_1 E^-$, with $\sign{(E^+)}=-\sign{(E^-)}$ as na\"ively expected, but importantly, with $\sign{(\partial_z E^+)} = \sign{(\partial_z E^-)}$. The order unity constant $\eta_1 \equiv |E^+| / |E^-|$ is set by the alignment of the optical trap to the center of the electrode cube. It is measured during calibration by considering the ratio of the single-electrode response to equal driving voltages on $z^+$ and $z^-$ with a fixed and known charge present on the MS, so that the monopole response is dominant. The position of the MS relative to the center of the electrode cube, $\vec{x}_0$, can then be inferred by comparison of the measured $\eta_1$ to the results from an FEA.

With the position of the MS known, we can define a second ratio of the gradients of $E^{\pm}$: $\eta_2 \equiv |\partial_z E^+| / |\partial_z E^-|$, implicitly evaluated at the position $\vec{x}_0$. This ratio is estimated from the FEA, making use of the measured position of the MS, derived as discussed above. Incidentally, for our system $\eta_1 \simeq \eta_2$ at the position of the optical trap, which we will just write as $\eta$ for brevity. With these tools in hand, we can construct a combined response parameter $A$, using the MS response amplitudes $F^{\pm} = (\vec{F}^{\pm} \cdot \vec{\hat{z}}) / \sin{(2 \pi f_0 t)}$ and the field amplitudes (dropping the explicit sinusoid):
\begin{align} \label{eq:a_parameter}
A &\equiv F^+ - \eta F^- \notag \\
&= q E^+ + p_{\mathrm{dc},z} \frac{\partial E^+}{\partial z} + p^+_{\mathrm{ac}} \frac{\partial E_{\mathrm{dc},z}}{\partial z} \notag \\
&\hspace*{0.5cm} - \eta \left( q E^- + p_{\mathrm{dc},z} \frac{\partial E^-}{\partial z} + p^-_{\mathrm{ac}} \frac{\partial E_{\mathrm{dc},z}}{\partial z} \right) \notag \\
&= 2 q E^+ + \left( p^+_{\mathrm{ac}} - \eta p^-_{\mathrm{ac}} \right) \frac{\partial E_{\mathrm{dc},z}}{\partial z}
\end{align}
\noindent where the combined monopole term has been simplified using $\sign{(E^+)}=-\sign{(E^-)}$ and $\eta = |E^+| / |E^-|$, and where the combined permanent dipole term has cancelled, by construction of $A$, since $\sign{(\partial_z E^+)}=\sign{(\partial_z E^-)}$. Assuming the field is uniform throughout the bulk of the MS, the induced dipole terms can be approximated as $p^{\pm}_{\mathrm{ac}} \simeq (4/3) \pi \epsilon_0 \chi r^3 E^{\pm}$ with $r$ being the radius of the MS. Then, the second term in Equation~\ref{eq:a_parameter} can be simplified to,
\begin{align} \label{eq:a_parameter_final}
A &\equiv F^+ - \eta F^- \notag \\
&= 2 q E^+ + \frac{8}{3} \pi \epsilon_0 \chi r^3 E^+ \frac{\partial E_{\mathrm{dc},z}}{\partial z}.
\end{align}

If $F^+$ and $F^-$ are added instead of subtracted,the component of $\vec{p}_{\mathrm{dc}}$ along the measurement axis can be estimated. Explicitly,
\begin{align} \label{eq:dc_dipole}
&F^+ + \eta F^- \notag \\
&\hspace*{0.3cm} = q E^+ + p_{\mathrm{dc},z} \frac{\partial E^+}{\partial z} + p^+_{\mathrm{ac}} \frac{\partial E_{\mathrm{dc},z}}{\partial z} \notag \\
&\hspace*{0.8cm} + \eta \left( q E^- + p_{\mathrm{dc},z} \frac{\partial E^-}{\partial z} + p^-_{\mathrm{ac}} \frac{\partial E_{\mathrm{dc},z}}{\partial z} \right) \notag \\
&\hspace*{0.3cm}= 2 p_{\mathrm{dc},z} \frac{\partial E^+}{\partial z}.
\end{align}

\subsection*{The Second Harmonic and the \ensuremath{B} Parameter}

Working in the regime of linear electrostatics, there is only a single forcing term at twice the fundamental driving frequency, as in Equation~\ref{eq:total_force}. Noting that $\vec{E}_{\mathrm{ac}} = E \sin{(2 \pi f_0 t)} \, \vec{\hat{z}}$ so that $\vec{p}_{\mathrm{ac}} = p_{\mathrm{ac}} \, \vec{\hat{z}}$, we consider this quantity for both $z^+$, $z^-$ electrodes,
\begin{align} \label{eq:second_harmonic}
\vec{p}_{\mathrm{ac}} \cdot \nabla \vec{E}_{\mathrm{ac}} &= p^{\pm}_{\mathrm{ac}} \frac{\partial E^{\pm}}{\partial z} \sin^2{(2 \pi f_0 t)}  \notag \\
&= \frac{p^{\pm}_{\mathrm{ac}}}{2} \frac{\partial E^{\pm}}{\partial z} \left\{ 1 - \cos{[2 \pi (2 f_0) t]} \right\},
\end{align}
\noindent where the constant term is usually ignored and we consider the term oscillating at $2 f_0$.

This expression suggests an interesting observable in that the response of the induced dipole at the second harmonic of the electrostatic drive is phase-shifted from the drive signal by $\pi/2$, in contrast to the monopole and permanent dipole terms which are expected to be in phase with the drive signal. The orthogonality of the induced dipole signal allows for individual probing and characterization, separate from the monopole and permanent dipole terms.

Let $G^{\pm} = (1/2) p^{\pm}_{\mathrm{ac}} \partial_z E^{\pm}$ be the amplitude of the MS response at the second harmonic $2 f_0$ (in equivalent force units) when the $z^{\pm}$, electrode is driven, respectively. Then we can construct a similar combined response parameter, $B$, for the second harmonic:
\begin{equation} \label{eq:b_parameter}
\begin{split}
B \equiv G^+ - \eta^2 G^- \simeq -\frac{1}{2} \left( p^+_{\mathrm{ac}} - \eta p^-_{\mathrm{ac}} \right) \frac{\partial E^+}{\partial z}.
\end{split}
\end{equation}

It can be seen that $B$ is proportional to the induced dipole term in the final expression for $A$. By combining Equations~\ref{eq:a_parameter_final} and \ref{eq:b_parameter}, this term can be eliminated in order to isolate the monopole response:
\begin{equation} \label{eq:monopole_final}
\begin{split}
2 q E^+ \simeq A + 2 B \frac{\partial_z E_{\mathrm{dc},z}}{\partial_z E^+}
\end{split}
\end{equation}

\section*{Upper Limits on neutrality of Matter and MCPs}

\subsection*{MCP search}

In order to convert the precision measurement of the charge of the three spheres to a limit on the amount of MCPs that bound to matter we introduce the following likelihood construction
\begin{equation} \label{eq:likelihood_one_sphere}
\Like_i^{(1)} (n_{\chi}, \epsilon) = \sum_{\rm N_i} {\rm Poisson}({\rm N_i};\, n_{\chi} {\rm N_{nucleons}})\times \Like_i^{(2)} (q_i= \epsilon {\rm N_i}),
\end{equation}
\noindent where $n_{\chi}$ is the abundance of MCPs per nucleon, ${\rm N_{nucleons}}$ is the number of nucleons in one MS, and ${\rm N_i}$ is the observed number MCPs in the sphere $i$. The charge of a sphere $i$ is simply the multiplication of ${\rm N_i}$ and $\epsilon$, the charge of an individual MCP. $\Like_i^{(2)}(q)$ is then use to model the charge response of the sphere $i$ in the following way:
\begin{equation} \label{eq:likelihood_mcp}
\Like_i^{(2)}(q) = \prod_{j} \frac{1}{\sqrt{2 \pi \sigma_{ij}^2}} \exp \bigg\{ \frac{-[A_{ij}/2 E_i^{{(+)}} - q_i)]^2}{2 \sigma_{ij}^2} \bigg\},
\end{equation}
\noindent where $j$ is running over all measured $A$s for a given sphere, and $\sigma_{ij}$ is estimated using the same fitting algorithm used to extract $A$, but on a sideband frequency separated from $f_0$ by 1\,Hz. 
Finally, the overall likelihood function is a multiplication of all three $\Like_i$ of the three MSs. 

\begin{figure}[!h]
    \includegraphics[width=1.\columnwidth]{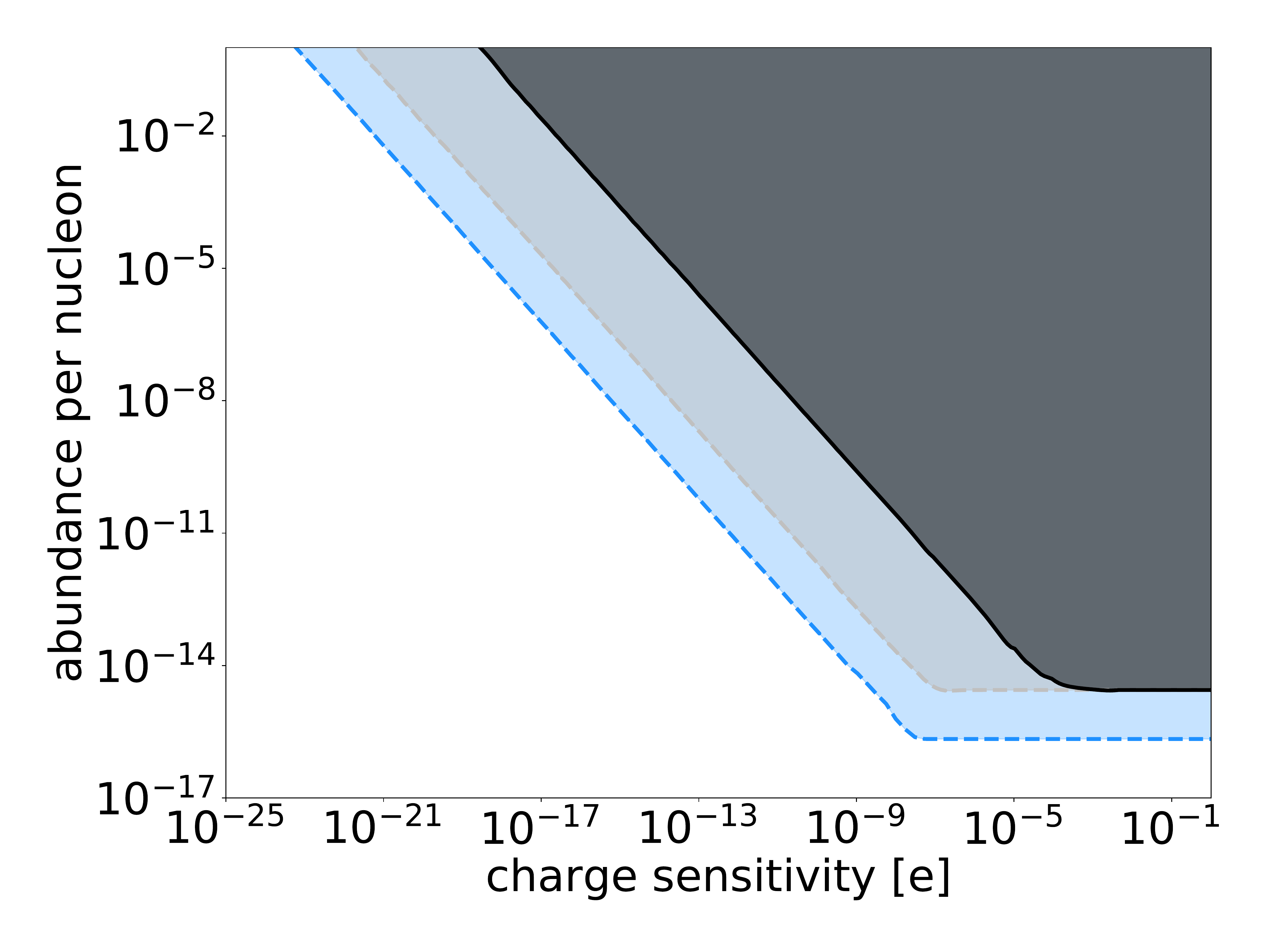}
    \caption{Translation of the measured charge sensitivity for all three MS combined into the parameter space of MCPs. For a given charge sensitivity the abundance has been profiled as described in the text. The black line represents our 95$\%$ exclusion limit given the measured overall charge on the MS presented here. For the projection used for the next iteration of the experiment two scenarios can be considered. The same sized spheres but with the improved SNR as described in the main text would yield the silver colored area. A different approach, where larger MS (r=15$\SI{}{\micro\meter}$), or equivalent several MS, are used for the same amount of time as in the text (t=92$\SI{}{hour}$) is indicated in blue. This scheme is sacrificing charge sensitivity for the sake of increased sensitivity to the neutrality of matter by the additional number of nucleons.)}
    \label{fig:mcp_limit}
\end{figure}

\begin{figure}[!h]
    \includegraphics[width=1.\columnwidth]{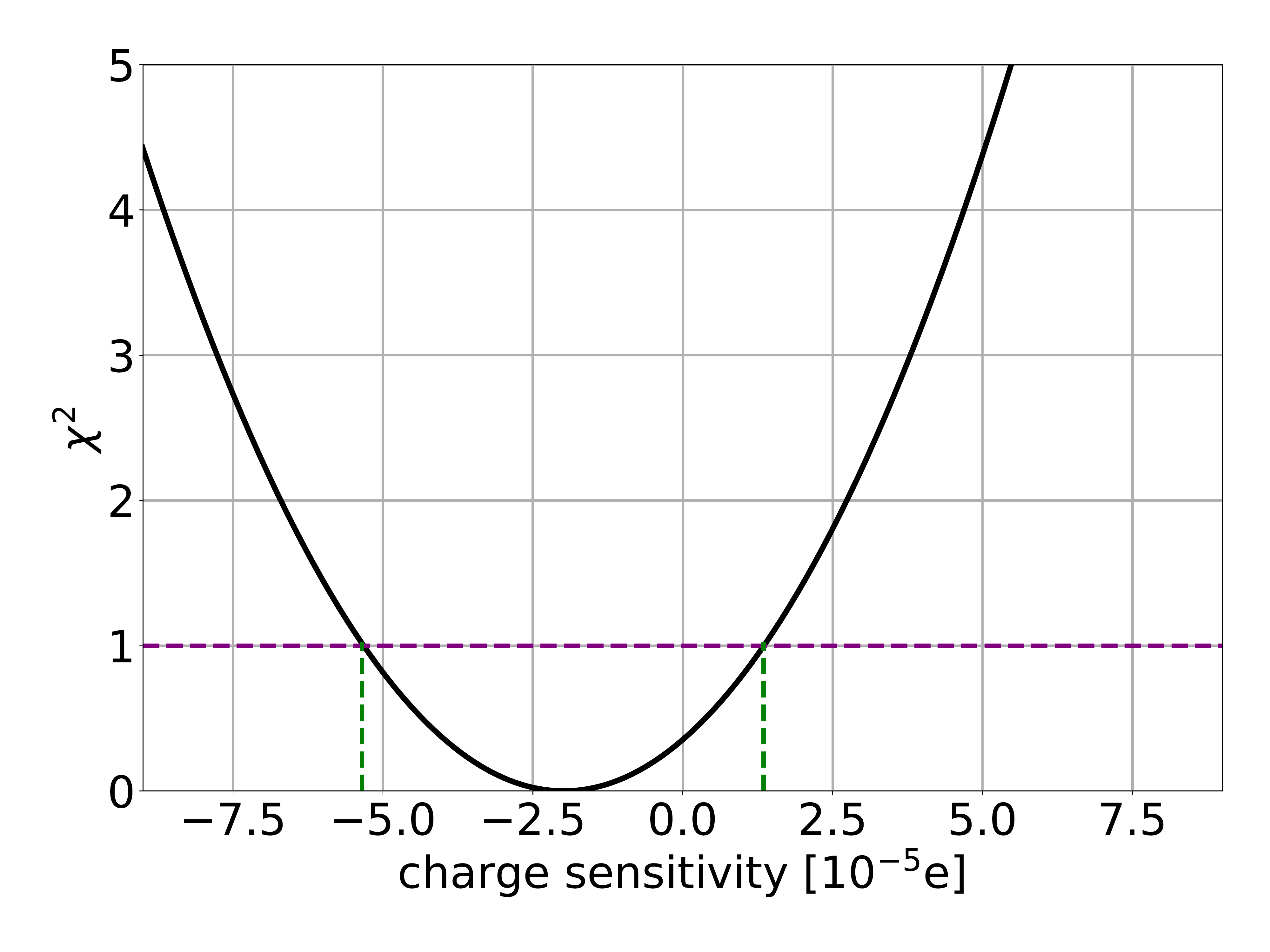}
    \caption{$\chi^2$-value profiled over the charge sensitivity. The corresponding line of $1\sigma$ is indicated dashed in purple and the intersection with the curve are indicated in dashed green lines. The result includes the zero signal value within this and the sensitivity is derived from the width between the two.}
    \label{fig:profile}
\end{figure}

Using this likelihood  and following the profile likelihood procedure described in~\cite{Cowan:2010} the data is tested and found to be in agreement with the no-signal (noise-only) hypothesis as show in \ref{fig:profile}, thus justifying our claim of a noise-limited measurement. Fig.~\ref{fig:mcp_limit} shows the achieved values in this work as well as extrapolations based on the improvements mentioned in the main text and the additional replacement of our existing sphere with larger ones (or the repetitive measurement of multiple spheres). 

\subsection*{Neutrality of Matter}

The combined measurement of the MSs global charge, $q$, from the main text can be translated to a limit on the equality of the electron and proton charge. The measurements of all three spheres are consistent with zero charge and their mass is \SI{420e-12}{\gram}. The combined limit results in $\SI{3.3e-5}{\elementarycharge}$ at the one-sigma level and can be translated to
$$(q_p-|q_e|)/e<2.6\times10^{-19},$$
\noindent where $q_p$ is the proton charge, and $q_e$ is the electron charge. Prior state of the art measurements of neutrality of matter have been able to achieve sensitives at the $10^{-21}$ level, and the understanding and precision achieved in this paper provides confidence that the next generation of neutrality of matter searches with levitated sphere will take the lead in this area (see Figure\,4 in the main text).

\bibliography{supplement}